# Visual-Informed Speech Enhancement Using Attention-Based Beamforming

Chihyun Liu, Jiaxuan Fan, Mingtung Sun, Michael Anthony, Mingsian R. Bai, *Senior Member*, IEEE, and Yu Tsao, *Senior Member*, IEEE

*Abstract*—Recent studies have demonstrated that incorporating auxiliary information, such as speaker voiceprint or visual cues, can substantially improve Speech Enhancement (SE) performance. However, single-channel methods often yield suboptimal results in low signal-to-noise ratio (SNR) conditions, when there is high reverberation, or in complex scenarios involving dynamic speakers, overlapping speech, or non-stationary noise. To address these issues, we propose a novel Visual-Informed Neural Beamforming Network (VI-NBFNet), which integrates microphone array signal processing and deep neural networks (DNNs) using multimodal input features. The proposed network leverages a pretrained visual speech recognition model to extract lip movements as input features, which serve for voice activity detection (VAD) and target speaker identification. The system is intended to handle both static and moving speakers by introducing a supervised end-to-end beamforming framework equipped with an attention mechanism. The experimental results demonstrated that the proposed audiovisual system has achieved better SE performance and robustness for both stationary and dynamic speaker scenarios, compared to several baseline methods.

*Index Terms*—multichannel speech enhancement, audiovisual, beamforming, deep learning, attention mechanism

## I. Introduction

SPEECH enhancement is a technique designed to enhance speech quality and intelligibility by mitigating background noise, interference, and reverberation. It plays a vital role in a variety of applications, including videoconferencing [1, 2], hearing aids [3, 4], voice assistants [5], and automatic speech recognition (ASR) systems [6–8]. Despite significant advances driven by deep learning, monaural SE approaches still face considerable challenges in complex acoustic environments, particularly in the presence of speech-like interference such as competing speakers or background television noise containing human dialog, where separating the target speech becomes inherently difficult.

To address this limitation, various auxiliary modalities have been explored to guide SE systems in focusing on the target speaker who is represented with speaker embeddings. Examples of such embeddings include i-vectors [9, 10], d-vectors [11, 12], and x-vectors [13], which have been widely adopted. These embeddings have demonstrated significant performance gains when incorporated into SE frameworks [14, 15]. However, these devices generally necessitate pre-enrollment and may prove ineffective in scenarios where competing speakers possess similar acoustic characteristics, such as those in the same gender.

In contrast to acoustic-based embeddings, visual cues such as facial and lip movements provide complementary cues that are inherently robust against acoustic interference. With lip-reading features or facial embeddings, SE systems can enhance the target speech signals more effectively, particularly in low-SNR and multi-speaker environments [16, 17]. Several approaches have been suggested to extract visual embeddings by cropping the image of the lip neighborhood for training and testing the SE models [18–21]. In order to reduce the computational complexity of image processing, strategies such as low-resolution lip region inputs [22] and bit-wise compression techniques [23, 24] have been reported. To expedite the extraction of visual features, task-specific pretrained networks have been introduced. For example, facial embeddings obtained from a pretrained face recognition model, FaceNet [25], are employed in [26] to capture facial expression features more accurately than raw image inputs. This representation has been demonstrated to mitigate speaker ambiguity, particularly in instances where speakers exhibit similar visual characteristics [27]. Furthermore, word-level lip-reading embeddings derived from a visual speech recognition (VSR) model [28] have been utilized in [29, 30] to inject linguistic information into the enhancement or separation pipeline. As stated in [31], the work introduces lip-reading embeddings extracted from a pretrained ASR model [32]. These embeddings enable effective detection of speech active and

This work was supported by the National Science and Technology Council (NSTC), Taiwan, under the project number 113-2221-E-007-057-MY3. (Corresponding author: Mingsian R. Bai).

Chihyun Liu is with the Department of Power Mechanical Engineering, National Tsing Hua University, Hsinchu, Taiwan (e-mail: joanna510119@gmail.com).

Jiaxuan Fan is with the Department of Electrical Engineering, National Tsing Hua University, Hsinchu, Taiwan (e-mail: frita7488@gmail.com).

Mingtung Sun is with the Department of Electrical Engineering, National Tsing Hua University, Hsinchu, Taiwan (e-mail: sunmtn0923@gmail.com).

Michael Anthony is with the Department of Electrical Engineering, National Tsing Hua University, Hsinchu, Taiwan (e-mail: michaelzhang220@gmail.com).

Mingsian R. Bai is with the Department of Power Mechanical Engineering and Electrical Engineering, National Tsing Hua University, Hsinchu, Taiwan (e-mail: msbai@pme.nthu.edu.tw).

Yu Tsao is with the Research Center for Information Technology Innovation, Academia Sinica, Taiwan (e-mail: minwelltsao@gmail.com).

silent regions and enhance intelligibility under extremely noisy conditions.

Depending on the features used, SE in general falls into three categories: Audio-Informed SE (AI-SE), Visual-Informed SE (VI-SE), and Audio-Visual-Informed SE (AVI-SE). Compared to models without any auxiliary information, VI-SE brings significant performance improvements [33], which is the main focus of this study. Although AVI-SE has been shown to outperform VI-SE in some studies [34], the performance gains are often marginal compared to the increased computational complexity and practical limitations introduced by the need for speaker voiceprint enrollment. In contrast, VI-SE has exhibited significant enhancements over AI-SE with minimal computational overhead. Consequently, this paper adopts the VI-SE approach as a lightweight solution for real-world applications.

While several state-of-the-art single-channel end-to-end VI-SE methods have been proposed [35–37], the absence of spatial information in such systems often leads to audible speech distortion, particularly for overlapping speech signals. Consequently, multichannel VI-SE frameworks that exploited spatial cues, such as inter-channel phase difference (IPD) and direction of arrival (DOA), were developed [21, 34]. Some multichannel techniques seek to estimate the parameters for beamformers directly from the audiovisual inputs [20, 38]. However, the efficacy of these methods is limited to stationary speakers and is not always applicable to nonstationary speakers.

To address the aforementioned limitations, we propose a novel framework, the Visual-Informed Neural Beamforming Network (VI-NBFNet), which leverages lip-reading features as auxiliary visual cues, in conjunction with microphone array beamforming techniques. The architecture jointly learns audiovisual features, mask representations, and spatial information in an end-to-end manner, with supervision directly at the beamforming output. In contrast to numerous two-stage data-driven systems, our approach eliminates the requirement for ground-truth mask supervision and facilitates unified optimization for all modules within the network. In this paper, we introduce an end-to-end system combining a time-varying Spatial Covariance Matrix (SCM) estimation model based on an attention mechanism [39] alongside a complementary visual-informed DeepFilter [40] as postfilter for SE.

The enhancement performance is evaluated using the following objective tests and performance metrics: Perceptual Evaluation of Speech Quality (PESQ) [41], Short-Time Objective Intelligibility (STOI) [42], and Deep Noise Suppression Mean Opinion Score (DNSMOS) [43]. A single-channel approach, modified from improved Lite Audio-Visual Speech Enhancement (iLAVSE) [24] and hereafter referred to as Visual-Informed Single-Channel Speech Enhancement (VI-SSE), is employed as one of the baselines in the evaluation. Furthermore, a mask-based beamformer, Visual-Informed Multichannel Speech Enhancement (VI-MSE) [44], [45], is adopted as another baseline. In this approach, time-frequency (T-F) masks are estimated using neural networks to compute SCMs for the beamforming weights. The final baseline, Visual-Informed Self-Attention Beamformer (VI-SA-BF), is an attention-based network that estimates time-varying SCMs directly from the instantaneous SCMs (ISCMs) [46].

The main contributions of this paper can be summarized as follows. First, we integrate lip movement features with audio features for SE networks within a multimodal framework. This integration has been demonstrated to have a substantial impact on the quality and intelligibility of speech, particularly in circumstances where acoustic conditions are unfavorable. Secondly, we propose a novel VI-NBFNet based on a hybrid architecture that integrates beamforming with deep learning to effectively suppress noise and interference while minimizing speech distortion. Finally, the proposed system employs a joint-trained attention network to learn spatial information in an end-to-end manner. This allows the model to estimate dynamic beamforming weights for the moving target speaker without an additional head tracker.

The remainder of this paper is organized as follows. Section II presents the problem formulation and the baseline systems. Section III introduces the proposed VI-NBFNet system. Section IV presents the experimental setups and the results. Conclusions and future work are given in Section V.

## II. PROBLEM FORMULATION AND BASELINE APPROACH

### A. Problem Formulation

Consider a scenario where a target speech signal and a directional interference are captured by $M$ microphones in a reverberant room. The signal received at the microphone can be written in the Short-Time Fourier Transform (STFT) domain as

$$Y^m(l,f) = A_s^m(f)S(l,f) + \sum_{n=1}^{N} A_n^m(f)I_n(l,f) + V^m(l,f), \quad (1)$$

where $l$ and $f$ denote the time frame index and the frequency bin index, respectively, $A_s^m$ denotes the Acoustic Transfer Function (ATF) of the $m$th microphone ($m = 1,2,…,M$) and the target speaker, $A_n^m$ denotes the ATF of the $m$th microphone and the $n$th interferer ($n = 1,2,…,N$), $S(l,f)$ denotes the target speaker signal, $I_n(l,f)$ denotes the $n$th interferer signal, and $V^m(l,f)$ denotes the $m$th sensor noise. The objective is to extract the target speaker's speech signal from the microphone signals in the presence of multiple directional interferers and additive sensor noise. It is assumed that the target speaker image is captured by the camera. This visual information facilitates the identification of the target speaker by the model.

### B. Baseline Methods

*1) Single-Channel Speech Enhancement System:* This baseline system is modified from the iLAVSE framework [24] and incorporates following three key features. First, rather than direct mapping, in which the network maps noisy signals directly to clean speech signals, a mask-based approach is employed. Secondly, low-dimensional lip embeddings extracted from a pretrained VSR model are employed in lieu of using the lip images. In the original iLAVSE, the image processing is conducted through a series of steps. The image data is reduced using color channel reduction, resolution downsampling, and quantization, followed by a Convolution Neural Network (CNN) based autoencoder to compress the visual inputs. In contrast, the current baseline utilizes embeddings that implicitly encode both phonetic content and speaker activities, providing richer and more informative visual cues than the original model. Next, the input to the model is extended from a context window to the full spectrogram. This allows the network to identify long-term temporal dependencies. Furthermore, the bi-directional Long-Short Term Memory (LSTM) previously employed in iLAVSE is substituted with a unidirectional LSTM to minimize latency for real-time processing. In summary, the complete baseline system, VI-SSE, as illustrated in Fig. 1(a), is a mask-based network that incorporates visual features extracted from a pretrained visual speech recognition VSR model as auxiliary information. In this system, a single T-F mask is estimated to process the spectrogram. The Mean Squared Error (MSE) between the enhanced and clean spectrograms is employed as the loss function for training the model. The network is denoted as modified iLAVSE (miLAVSE).

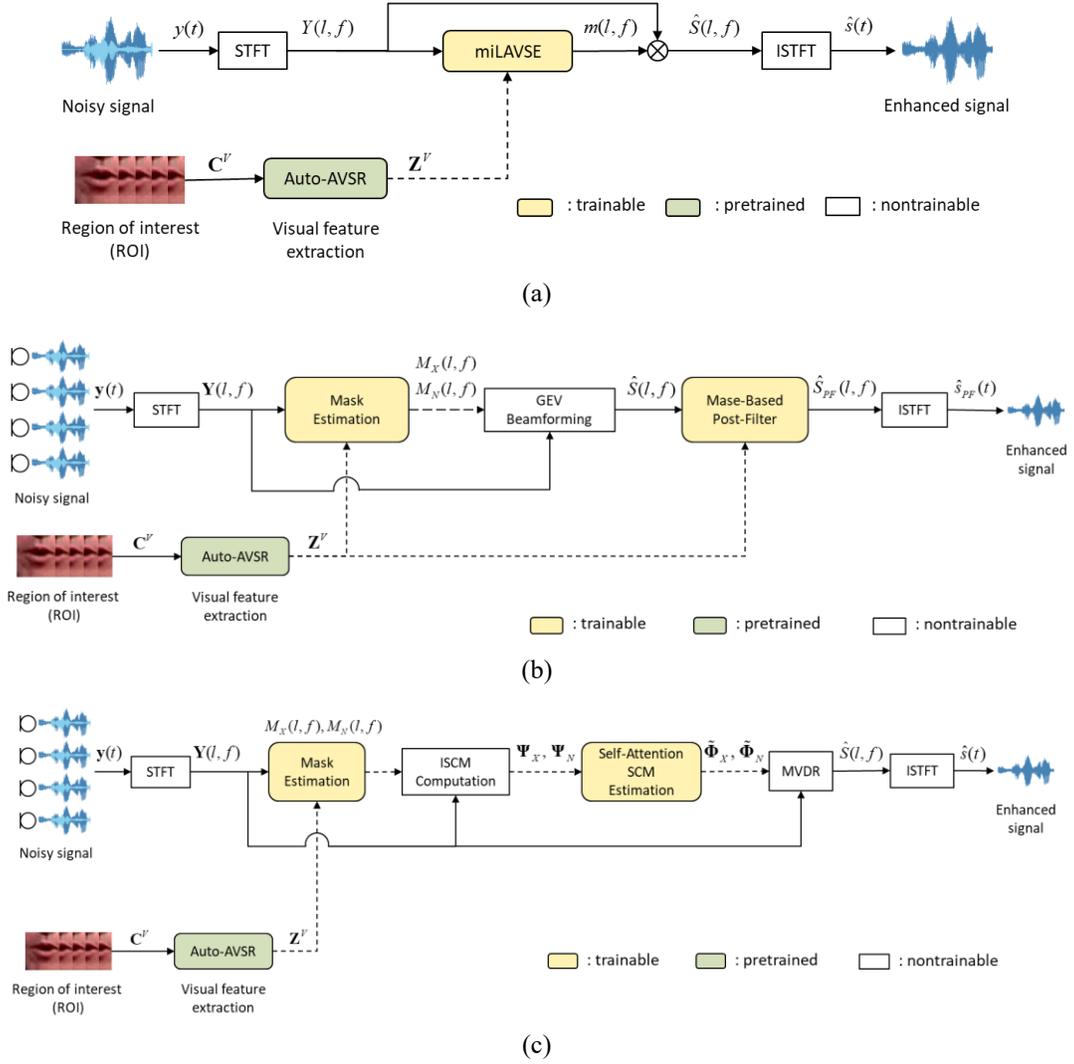

Fig. 1. Block diagram of the baseline methods. (a) VI-SSE system, (b) VI-MSE system, (c) VI-SA-BF system. Yellow and green blocks represent trainable and pretrained networks, while white blocks denote non-trainable modules.

2) Neural Network-Based Beamformer and Postfilter: The baseline system, VI-MSE, is illustrated in Fig. 1(b). The system is composed of three primary modules: a mask estimation module, a beamforming module, and a postfilter module. The mask estimation module employs a conventional pipeline that has been utilized in a recent neural beamformer [44], wherein the T-F masks for speech and noise are estimated. These masks are utilized to compute the SCMs required for the beamformer. The mask estimation network in VI-MSE is similar to that of miLAVSE, except that it predicts two Ideal Ratio Masks (IRMs) instead of one. The mask estimation network is depicted in Fig. 2. It is noteworthy that, while [44] adopts Ideal Binary Masks (IBMs) that utilize clean speech and noise as training targets, our study employs Wiener-Like Masks (WLMs) [47], which provide a smoother and more informative supervision signal. In the beamforming module, the time-invariant speech and noise SCMs are computed for each frequency bin using Eqs. (2) and (3). As suggested in [44], the generalized eigenvalue (GEV) beamformer combined with a Blind Analytical Normalization (BAN) postfilter is utilized to obtain the enhanced signals at the final stage.

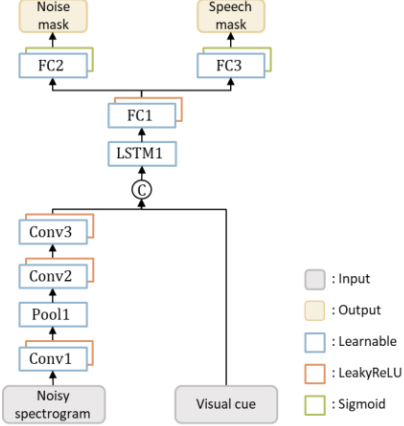

Fig. 2. The network architecture of the mask estimation network.

$$\mathbf{\Phi}_{XX}(f) = \frac{\sum_{l=1}^{T} M_X(l,f)\mathbf{Y}(l,f)\mathbf{Y}(l,f)^H}{\sum_{l=1}^{T} M_X(l,f)} \in \mathbb{C}^{M \times M} \quad (2)$$

$$\mathbf{\Phi}_{NN}(f) = \frac{\sum_{l=1}^{T} M_N(l,f)\mathbf{Y}(l,f)\mathbf{Y}(l,f)^H}{\sum_{l=1}^{T} M_N(l,f)} \in \mathbb{C}^{M \times M} \quad (3)$$

where $T$ denotes the number of total time frames, $F$ denotes the total number of frequency bins, $\mathbf{\Phi}_{XX}(f)$ and $\mathbf{\Phi}_{NN}(f)$ denote the time-invariant speech and noise SCMs at frequency bin $f$, $M_X(l,f)$ denotes predicted speech mask, and $M_N(l,f)$ denotes predicted noise mask. $(\cdot)^H$ denotes the Hermitian transpose. The beamformer, operating as a preprocessor, has demonstrated efficacy in suppressing noise sources that are spatially separable. However, in some cases, residual noise remains audible in the output. To address this issue, a learning-based postfilter is employed to further reduce the noise, especially for collocated sources. The postfilter network is similar to the miLAVSE network, but it accepts the beamformed spectrogram as the input. In this case, the clean speech spectrogram serves as the training target. The mask estimation network and the postfilter are trained with the MSE loss.

3) Self-Attention-Based Neural Beamformer: The baseline is based on the framework proposed in [46] for estimating the time-varying SCM. The original method utilizes a TasNet [48] to estimate T-F masks for single-speaker scenarios. The architecture is modified to accommodate multi-source scenarios through the integration of a visual-informed mask estimation module derived from the VI-MSE system. Specifically, the TasNet-based mask estimator is substituted with the miLAVSE-based mask estimator to incorporate visual cues associated with the target speaker, which is crucial for accurate mask estimation when both target and interferers coexist. The implementation of the self-attention-based SCM estimation and the neural beamformer remains unaltered. The preceding architecture is referred to as the VI-SA-BF, as illustrated in Fig. 1(c).

III. PROPOSED METHOD

In this section, we propose a VI-NBFNet to address challenging SE tasks. The proposed system adopts a hybrid architecture that integrates microphone array signal processing (MASP) with DNNs. A key feature of the network is its end-to-end joint learning of audiovisual and spatial information, which enables the prediction of weights necessary for time-varying SCM estimation. A schematic overview of the proposed system is shown in Fig. 3. The system consists of four main components: the audiovisual encoder, mask decoder, spatially aware decoder, and SCM computation module. Despite the architecture of a modular design, the components are jointly optimized end-to-end, allowing the network to effectively learn from the spatial and spectral features.

A. Visual feature extraction

Lip movement is considered a highly reliable VAD and a word-relevant feature for ASR. Similarly, lip reading can be exploited as a powerful tool for SE tasks. The pretrained Audio-Visual Speech Recognition (AVSR) model (auto-AVSR) [49], is employed to extract the visual features for the proposed system and all baselines. The front end of auto-AVSR is composed of an ASR encoder and a VSR encoder. The input to the ASR encoder is the noisy speech signals. The input to the VSR encoder is the corresponding cropped lip imagery $\mathbf{C}^{Visual} \in \mathbb{R}^{T^{Visual} \times 96 \times 96}$ of the target speaker, with $T^{Visual}$ representing the number of image frames. The outputs of the audio and video encoders are then integrated through the application of a Multi-Layer Perceptron (MLP). The VSR encoder is responsible for generating the lip-reading features $\mathbf{Z}^{Visual} \in \mathbb{R}^{T^{Visual} \times 512}$. These latent features are not only relevant to voice activity but also to word embedding associated with speech articulation, which is widely used in ASR. The output of the MLP feeds into the back end, which comprises a projection layer and a transformer decoder. In the proposed system, the aforementioned latent features are utilized for the SE tasks. Given that the video stream operates at a frame rate of 25 frames per second (fps), while the audio stream operates at a frame rate of 100 fps, temporal alignment is necessary between the two modalities. Thus, each video frame is repeated four times to align with the audio stream to ensure that the visual features and noisy audio spectrograms are synchronized along the time axis. This step is crucial for the subsequent stages of multimodal fusion.

B. Audiovisual Encoder

The VI-NBFNet employs a similar audiovisual fusion strategy to that of iLAVSE, where audio and visual features are concatenated and fed into an LSTM layer to detect temporal dependencies. However, the configuration of audio encoders differs between the two models. Although iLAVSE utilizes a straightforward three-layer CNN designed for single-channel mask estimation, VI-NBFNet employs a lightweight variant of MobileNetV2 [50] to more effectively process multichannel inputs and generate more expressive representations. This approach is motivated by the necessity to capture more salient features from microphone array signals, which are vital for estimating the weights required in SCM computation for beamforming. Despite the fact that MobileNetV2 was not originally designed for spatial modeling, its use of pointwise and depthwise separable convolutions provides a trade-off between computational complexity and representation effectiveness. This architecture serves as the backbone of our audio encoder, as illustrated in Fig. 5. Specifically, the initial pointwise (1×1) convolutions linearly combine the stacked input channels, thereby enabling the network to learn inter-microphone correlations. Then, depthwise (3×3) convolutions are employed to extract local T-F patterns in each channel. Although this network does not explicitly encode spatial magnitude and phase difference of the audio signals, the implicit encoding of spatial information is achieved through the stacking of the multichannel real and imaginary parts along the channel axis. The learned representations have been shown to facilitate downstream modules, such as the spatially aware decoder and attention-based SCM estimation blocks, in more effective modeling of spatial characteristics.

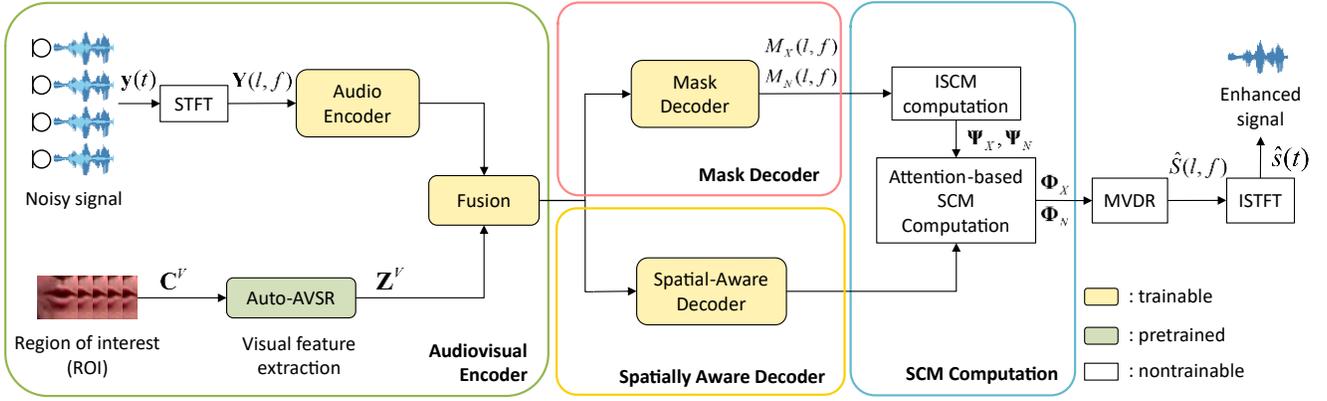

Fig. 3. Schematic diagram of VI-NBFNet. Yellow and green blocks represent trainable and pretrained networks, while white blocks denote non-trainable modules.

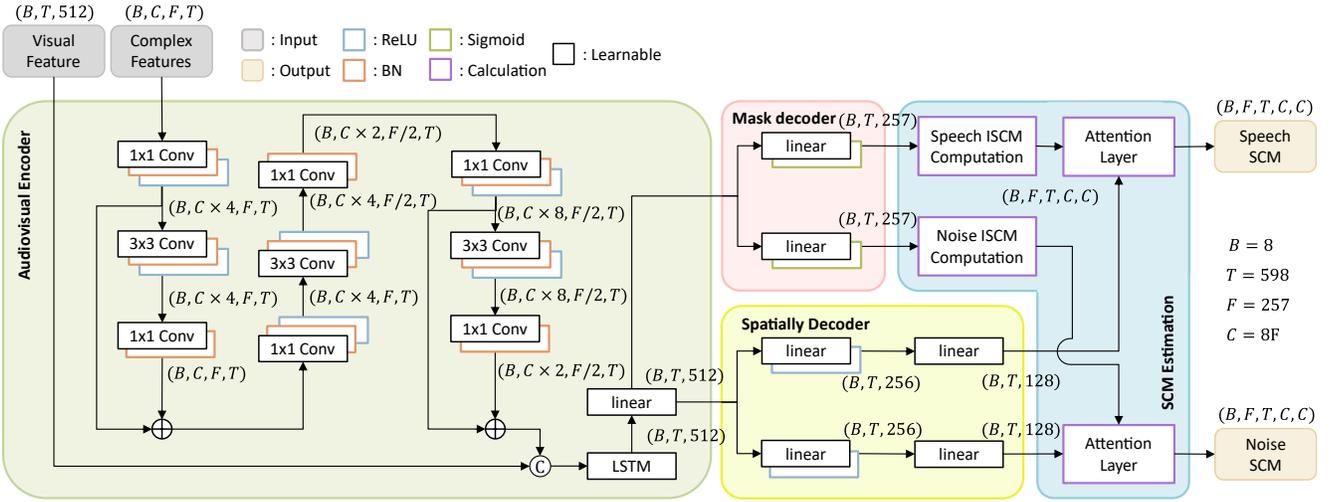

Fig. 4. The network architecture and tensor parameters of VI-NBFNet.

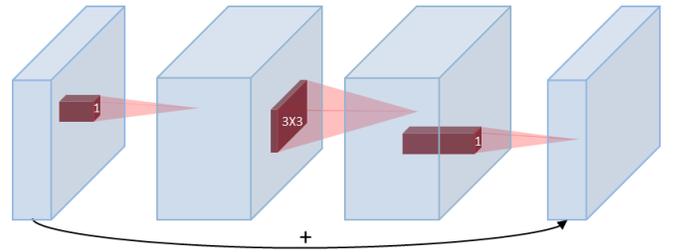

Fig. 5. The Inverted Residual Structure of MobileNetV2.

### C. Mask Decoder and Spatially Aware Decoder

The present network is distinct from the VI-SA-BF system in two primary respects. Firstly, the entire framework is trained end-to-end, as opposed to adopting a two-stage pipeline that separates mask estimation from the SCM computation. Secondly, rather than employing the instantaneous spatial covariance matrix (ISCM) as both the query and key inputs to the attention mechanism, we introduce a dedicated spatially aware decoder that generates independent feature representations for attention weight estimation. In both approaches, however, the ISCM is used as the value input to the attention layer in line with the SCM formulation suggested in [46] as follows:

$$\tilde{\mathbf{\Phi}}_X(t,f) = \sum_{t'=1}^{T} c(t,t')\mathbf{\Psi}_X(t',f) \in \mathbb{C}^{M\times M}, \quad (4)$$

$$\tilde{\mathbf{\Phi}}_N(t,f) = \sum_{t'=1}^{T} c(t,t')\mathbf{\Psi}_N(t',f) \in \mathbb{C}^{M\times M}, \quad (5)$$

where $\tilde{\mathbf{\Phi}}_X(t,f)$ and $\tilde{\mathbf{\Phi}}_N(t,f)$ denote the time-varying speech and noise SCMs at the T-F bin $(t,f)$. $c(t,t')$ denotes the weighting coefficients associated with the contribution of the frame $t'$ when estimating the SCM at the current frame $t$. The ISCMs at frame $t'$ for the same frequency bin $f$, $\mathbf{\Psi}_X(t',f)$ and $\mathbf{\Psi}_N(t',f)$, are computed using

$$\mathbf{\Psi}_X(t,f) = M_X(t,f)\mathbf{Y}(t,f)\mathbf{Y}^H(t,f) \in \mathbb{C}^{M\times M} \quad (6)$$

$$\mathbf{\Psi}_N(t,f) = M_N(t,f)\mathbf{Y}(t,f)\mathbf{Y}^H(t,f) \in \mathbb{C}^{M\times M} \quad (7)$$

The preceding ISCMs are obtained by applying the speech and noise masks to the multichannel input. This process necessitates a mask decoder to generate the respective masks. The mask decoder is based on the mask estimation network used in the miLAVSE system. Specifically, audiovisual fusion is achieved through an LSTM layer in conjunction with a shared linear projection. Then, the output data is directed to two parallel linear layers and sigmoid activation to predict the speech and noise masks. In order to prevent the attention weight estimation from being impacted by the mask learning process, a spatially aware decoder is employed as a separate module. This decoder is implemented as a lightweight MLP consisting of two linear layers with a ReLU activation in between, projecting the fused audiovisual features from 512 to 128 dimensions. That is, (B, T, 512) → (B, T, 256) → (B, T, 128). By transforming the shared representations into a compact feature space, the spatially aware decoder provides independent spatial cues to guide the attention mechanism to capture temporal dependencies. This procedure is imperative for precise spatial covariance estimation.

*D. Computation of Attention-Based SCM*

The computation of attention-based SCMs may bear a resemblance to the method outlined in [46]. However, upon closer examination, it becomes evident that these methods differ significantly in terms of complexity and learnability. This discrepancy is evident in the parameter sizes and Multiply-Accumulate (MAC) operations summarized in Table I. The time-varying SCM is estimated using a stacked self-attention network, as described in [46]. This network consists of layers analogous to Transformer encoders, incorporating not only an attention mechanism but also additional feedforward sublayers. In contrast, the SCM computation block in VI-NBFNet contains no learnable unit. However, its physically interpretable structure, in conjunction with end-to-end training that is supervised by the beamforming output, allows the model to learn effectively the spatial representations. The computation is based on an attention mechanism, in which scaled dot-product operations are performed with the query and key, followed by Softmax activation to generate the attention weights. The instantaneous ISCMs are used as the value input, and the attention weights are applied to the summation across time frames as follows:

$$\mathbf{A} = \text{Weight}(\mathbf{Q},\mathbf{K}) = \text{Softmax}(\frac{\mathbf{Q}\mathbf{K}^T}{\sqrt{d_k}}), \quad (8)$$

$$\mathbf{Z} = \text{Att}(\mathbf{A},\mathbf{V}) = \mathbf{A}\mathbf{V}, \quad (9)$$

where $\mathbf{Q} \in \mathbb{R}^{T\times d_k}$ and $\mathbf{K} \in \mathbb{R}^{T\times d_k}$ denote the query and key matrices, respectively, which are obtained from the spatially aware decoder, $d_k$ is the feature dimension of the query and key, the superscript $(\cdot)^T$ indicates transpose, and $\mathbf{V} \in \mathbb{R}^{T\times M\times M}$ denotes the value corresponding to the ISCM. This process yields the time-varying SCM, as illustrated in Fig. 6. The computation of the attention weights $\mathbf{A} \in \mathbb{R}^{T\times T}$ is conducted in accordance with Eq. (8), and the final output $\mathbf{Z}$ is obtained using Eq. (9). When $\mathbf{V}$ represents ISCMs, the resulting $\mathbf{Z}$ becomes the time-varying SCM. The formulation above is mathematically equivalent to the formulations in Eqs. (4) and (5), thereby rendering the attention mechanism interpretable.

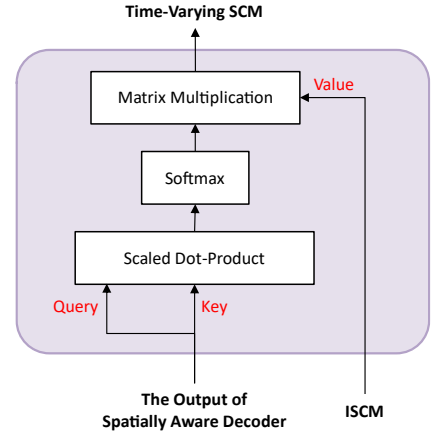

Fig. 6. The attention layer for the estimation of time-varying SCM.

TABLE I
PARAMETER SIZE AND MACs OF VI-NBFNET AND SCM ESTIMATION NETWORK OF VI-SA-BF.

|  | VI-NBFNet | VI-SA-BF |
|---|---|---|
| Parameters (M) | 7.15 | 14.7 |
| MACs (G/s) | 0.4 | 0.6 |

*E. Minimum-Variance Distortionless Response (MVDR) Beamformer*

Once the time-varying SCMs are obtained, an MVDR beamformer is implemented [49].

$$\mathbf{w}_{MVDR}(t,f) = \frac{\tilde{\mathbf{\Phi}}_N(t,f)^{-1}\tilde{\mathbf{\Phi}}_X(t,f)}{\text{Tr}(\tilde{\mathbf{\Phi}}_N(t,f)^{-1}\tilde{\mathbf{\Phi}}_X(t,f))}\mathbf{u}, \quad (10)$$

where Tr(·) denotes the matrix trace operator. $\mathbf{u} \in \mathbb{R}^M$ is a one-hot vector associated with the reference microphone. It follows that the beamforming output is given by

$$\hat{S}(t,f) = \mathbf{w}_{MVDR}^H(t,f)\mathbf{Y}(t,f), \quad (11)$$

where $\hat{S}(t,f)$ denotes the beamformed signal in the STFT domain, which serves as the final output of the model. Although both our method and VI-SA-BF utilize the MVDR beamforming result as the network output, the adopted loss functions differ. In our framework, the objective function comprises a MSE loss computed in the T-F domain and an SNR loss computed in the time domain. The formulation is as follows:

$$loss = \frac{100}{FT}\sum_{f=1}^{F}\sum_{l=1}^{T}\left(|S(l,f)| - |\hat{S}(l,f)|\right)^2 + \frac{1}{T}\sum_{n=1}^{T}\frac{|s(n)|^2}{|s(n)-\hat{s}(n)|^2}, \quad (12)$$

where $S(t,f)$ and $s(n)$ represent the ground truth (clean) signals at the reference microphone in the T-F and time domains, respectively. Similarly, $\hat{S}(t,f)$ and $\hat{s}(n)$ represent the corresponding clean signals estimated by beamforming. $T$ is the number of time frames in the T-F domain, and T represents the number of time samples in the time domain.

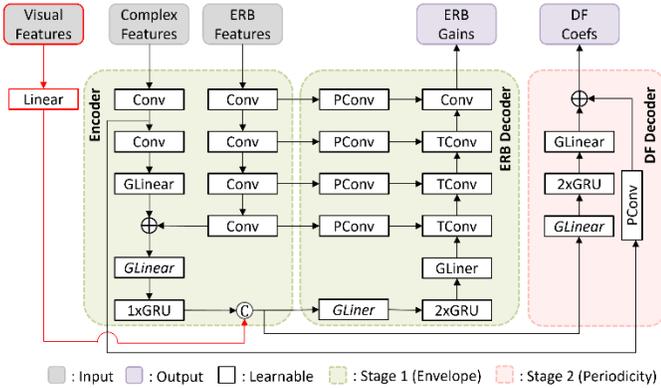

Fig. 7. The network architecture of Visual-Informed DeepFilter-based post-filter.

*F. Visual-Informed DeepFilterNet*

In addition to the VI-NBFNet, an optional postfilter termed Visual-Informed DeepFilter (VIDF) is proposed to further enhance the noise reduction performance, albeit at the expense of a slight increase in computational complexity. The architectural design of VIDF is illustrated in Fig. 7. The postfilter design is inspired by DeepFilterNet2 [40], a lightweight DNN that has been demonstrated to be capable of suppressing the residual noise present at the output end of a DNN. The original DeepFilterNet2 is composed of an encoder and a two-stage decoder, which includes an Equivalent Rectangular Bandwidth (ERB) gain prediction stage and a deep filtering stage. In this study, we augment the input features with visual cues to improve SE performance, particularly in scenarios involving competing speakers. A similar fusion strategy is adopted, analogous to miLAVSE, in which audio and visual features are concatenated after the audio encoder and passed through an LSTM layer to capture temporal dependencies. The encoder output is then concatenated with the lip-reading visual features, and the fused representations are fed into the ERB decoder and the deep filter decoder. Consequently, both decoders of VIDF incorporate GRU layers, which operate in a manner analogous to the LSTM in miLAVSE, intended for modeling the temporal dynamics of speech guided by visual cues.

IV. EXPERIMENTS

*A. Dataset and setup*

In this study, the Lip Reading Sentences 3 -TED (LRS3-TED) dataset [52], consisting of 400 hours of TED and TEDx talks, was employed to generate the training and test data. In the training and testing phases, a total of 800 speakers were selected from this corpus, while 50 speakers were selected as the test set. To create diverse interference conditions, multiple datasets were incorporated. The LibriSpeech corpus [53], comprising 1000 hours of English read speech from public domain audiobooks, was utilized to represent competing speakers. Music-related interference was also introduced using samples from the Free Music Archive (FMA) dataset [54]. For stationary noises typically encountered in office environments, selected classes from the Microsoft Scalable Noisy Speech Dataset (MS-SNSD) dataset [55], including vacuum cleaner, air conditioner, and copy machine, were utilized. All audio signals were sampled at 16 kHz.

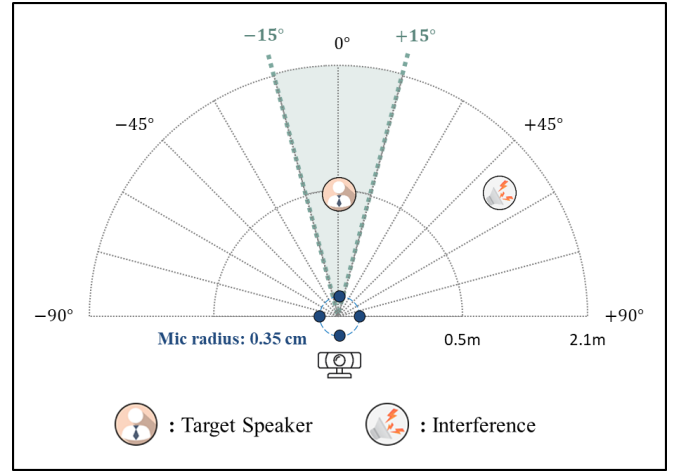

Fig. 8. Illustration of the experimental settings in the training phase.

In the training phase, a four-microphone uniform circular array with a radius of 3.5 cm, ReSpeaker® Mic Array v2.0, was employed. To simulate realistic multi-source environments, a total of 30,000 training samples and 9,000 validation samples were generated. Each sample comprised a six-second segment of clean speech, mixed with various types of interference. The target speech was combined with all interference at Signal-to-Interference Ratios (SIRs) of –10, –5, 0, 5, 10, and 15 dB. In addition, Gaussian noise with a SNR randomly selected between 30 and 40 dB was added independently to each channel to emulate sensor noise present in real microphone recordings. All sources, including the target speaker and interferers, were positioned on a semicircular arc in front of the array, with radial distances of 0.5 to 2.1 m from the array center. The target speaker's azimuthal angle was constrained in the range of –15° to 15° to ensure visibility within the camera's effective angle of view. The target was closer to the center of the array than the interference in static conditions. However, the target speaker moved along a straight line within the same angular range in dynamic conditions. Despite the variability in movement speed across different samples, a constant velocity was maintained throughout the six-second utterances in each sample to simulate speaker motion in a realistic manner. The experimental setup is illustrated in Fig. 8. Room impulse responses (RIRs) with reverberation time (RT60) ranging from 200 to 500 ms were generated using the image source method [56]. These RIRs were then convolved with clean source and interference signals to create multichannel reverberant mixtures.

The real-data recording using the microphone array used in training was conducted in a conference room located in Engineering Building I at National Tsing Hua University in Taiwan, as illustrated in Fig. 9. The $7.0 \text{ m} \times 5.3 \text{ m} \times 2.7 \text{ m}$ room has RT60 of approximately 430 ms. The video was recorded using an Apple® iPad Air 5. This configuration facilitated the concurrent acquisition of audiovisual data within a realistic reverberant environment. The target speech was recorded from a human speaker or from playback through a loudspeaker. The interference signals comprised loudspeaker playback of non-target speech, and separately, ambient noise from the air-conditioning system.

The simulated test data followed the same configuration as the training setup, except that all utterances were drawn from unseen speakers. A total of 1,800 samples were generated for testing, comprising 900 samples with a target speaker and 900 samples with a moving target speaker. These two datasets were used to evaluate the system's performance under static and dynamic conditions, respectively.

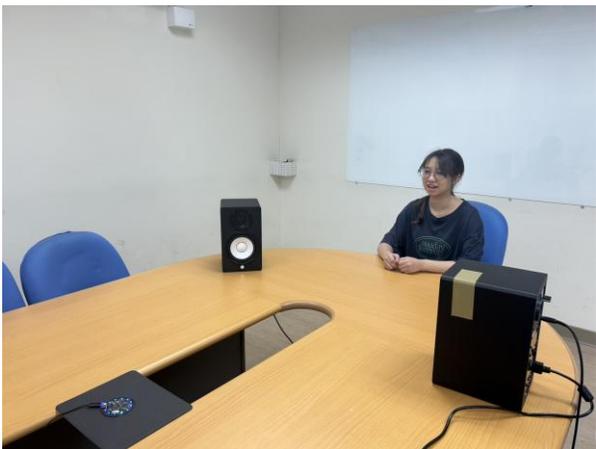

Fig. 9. Experimental environment for audiovisual recording.

The proposed models and baselines were trained using a frame length of 25 ms and a time step of 10 ms, in conjunction with the Adam optimizer. The 512-point Fast Fourier Transform (FFT) was employed.

*B. Evaluation metrics*

To assess the enhancement performance of the proposed systems, three objective metrics were adopted. The PESQ and STOI metrics were employed to assess speech quality and intelligibility in the context of intrusive evaluation. Conversely, for non-intrusive assessment, the DNSMOS metric [43], trained on P.835 human ratings, was employed as a speech quality index. DNSMOS has three performance indices: speech quality (SIG), background noise quality (BAK), and overall quality (OVRL). In general, SIG and BAK is intended to indicate the trade-off between speech distortion and noise reduction.

*C. Simulation Results*

A comprehensive evaluation was conducted to validate the proposed systems. The baseline systems and the proposed method were categorized into two groups according to their architectural characteristics and output formats. Since the proposed VI-NBFNet produces a beamforming signal, the first group of baselines that we select include VI-MSE (BF) and VI-SA-BF, both of which also produce beamforming outputs. The second group of baselines, VI-SSE and VI-MSE (BF+PF), includes an additional postfilter for benchmarking the proposed network when it is cascaded with a DF as a postfilter. Having established these baseline groups, we then conducted a series of additional analyses to comprehensively evaluate the proposed VI-NBFNet. The evaluation covers five aspects: (1) a comparative analysis of multichannel audiovisual systems equipped with beamforming technology and their single-channel counterparts, (2) examination of the effects of using a postfilter, (3) an ablation study to validate the contribution of the proposed joint loss function defined in Eq. (12), (4) visualization of the learned attention weights to interpret the behavior of the Spatial Awareness Decoder (SAD), and (5) robustness evaluation under various visual degradation conditions. These experiments collectively provide a comprehensive understanding of the performance, interpretability, and resilience of the proposed audiovisual beamforming framework.

TABLE II
AVERAGE PESQ, STOI, AND DNSMOS FOR PROPOSED METHOD AND BEAMFORMING-BASED BASELINES ACROSS SIRS OF -10, -5, 0, 5, 10, AND 15 DB.

|  | Stationary target speaker | | | | |
| --- | --- | --- | --- | --- | --- |
|  | PESQ | STOI | BAK | SIG | OVRL |
| Noisy | 1.342 | 0.659 | 1.391 | 1.713 | 1.308 |
| VI-MSE(BF) | 1.588 | 0.768 | 1.749 | 2.144 | 1.573 |
| VI-SA-BF | 1.559 | 0.768 | 3.013 | 2.254 | 1.732 |
| VI-NBFNet | **1.954** | **0.835** | **3.735** | **2.570** | **2.177** |
|  | Moving target speaker | | | | |
|  | PESQ | STOI | BAK | SIG | OVRL |
| Noisy | 1.334 | 0.656 | 1.383 | 1.699 | 1.303 |
| VI-MSE(BF) | 1.483 | 0.736 | 1.658 | 1.946 | 1.484 |
| VI-SA-BF | 1.554 | 0.764 | 3.027 | 2.212 | 1.701 |
| VI-NBFNet | **2.026** | **0.849** | **3.749** | **2.569** | **2.183** |

*1) Comparison of VI-NBFNet, VI-MSE (BF), and VI-SA-BF:* The results are presented in Table II. For moving speaker scenarios, VI-SA-BF and VI-NBFNet obtain higher average scores than VI-MSE (BF) in all evaluation metrics (PESQ, STOI, and DNSMOS). This outcome demonstrates the advantage of using time-varying SCMs over time-invariant SCMs for beamforming, especially with respect to the

DNSMOS metrics. However, for stationary speaker scenarios, VI-SA-BF shows slightly lower PESQ and STOI scores than VI-MSE (BF). This difference can be attributed to inaccurate time-varying SCM estimation at low SIR levels, which can introduce mild distortions compared to clean speech.

Notably, the proposed VI-NBFNet outperforms all beamforming-based baselines in stationary and moving speaker scenarios with regard to all evaluation metrics and SIR levels, as summarized in Table II. Moreover, the improvements in DNSMOS, especially in SIG and BAK, indicate that the proposed model yields superior enhancement in target speech quality and interference suppression. It is worth noting that a significant improvement in performance can be observed in both PESQ and STOI at low SIR levels, e.g., –10 and –5 dB. This finding indicates that the joint learning strategy employed by VI-NBFNet demonstrates satisfactory generalization and robustness in adverse acoustic conditions, in contrast to VI-SA-BF which relies on a two-stage mask estimation and self-attention-based beamforming.

2) Comparison of VI-NBFNet, VI-NBFNet-DF, VI-SSE, and VI-MSE (BF+PF): The results are presented in Table III. An examination of the summary reveals that all multichannel approaches demonstrate superior performance in comparison to the single-channel baseline (VI-SSE) across all performance metrics, particularly in PESQ, STOI, and SIG. This observation underscores the significance of spatial cues, which are absent in single-channel systems and render them more susceptible to speech distortion.

TABLE III
AVERAGE PESQ, STOI, AND DNSMOS FOR PROPOSED METHOD, VI-SSE AND VI-MSE ACROSS SIRs OF -10, -5, 0, 5, 10, AND 15 dB.

|  | Stationary target speaker | | | | |
|---|---|---|---|---|---|
|  | PESQ | STOI | BAK | SIG | OVRL |
| Noisy | 1.342 | 0.659 | 1.391 | 1.713 | 1.308 |
| VI-SSE | 1.558 | 0.750 | 3.129 | 2.448 | 2.014 |
| VI-MSE (BF+PF) | 1.798 | 0.811 | 3.355 | 2.748 | 2.301 |
| VI-NBFNet | 1.954 | 0.835 | **3.735** | 2.570 | 2.177 |
| VI-NBFNet-DF | **2.015** | **0.841** | 3.474 | **2.813** | **2.399** |
|  | Moving target speaker | | | | |
|  | PESQ | STOI | BAK | SIG | OVRL |
| Noisy | 1.334 | 0.656 | 1.383 | 1.699 | 1.303 |
| VI-SSE | 1.553 | 0.746 | 3.146 | 2.439 | 2.010 |
| VI-MSE (BF+PF) | 1.651 | 0.783 | 3.251 | 2.571 | 2.134 |
| VI-NBFNet | 2.026 | 0.849 | **3.749** | 2.569 | 2.183 |
| VI-NBFNet-DF | **2.088** | **0.855** | 3.597 | **2.838** | **2.438** |

Among the multichannel methods, VI-MSE (BF+PF) has been demonstrated to achieve higher SIG scores than VI-NBFNet, thereby indicating that the combination of beamforming and a postfilter is highly effective in improving speech quality. This advantage is particularly evident in scenarios involving stationary speakers, for which fixed beamforming is well suited. Nevertheless, the VI-NBFNet model, when implemented without a postfilter, demonstrates superior performance in comparison to the VI-MSE (BF+PF) model across most of the evaluated metrics. VI-NBFNet has achieved the highest BAK scores, demonstrating strong noise suppression capability. While VI-NBFNet yields slightly lower SIG in the stationary speaker condition, it demonstrates consistent enhancement performance and robustness to the moving target speaker.

Additionally, we evaluated the effect of applying a Visual-Informed DeepFilter-based postfilter following VI-NBFNet. As shown in Table III, the SIG score improves significantly, suggesting enhanced perceived speech quality. However, a slight decrease in BAK is observed, which may be due to residual artifacts introduced by the postfilter. In this particular instance, while the majority of performance metrics have demonstrated improvements, the postfilter has been observed to incur greater memory storage and computational overhead. As a result, this configuration was not included in the subsequent experiments.

3) Ablation study on loss function: To further validate the effectiveness of the proposed joint loss function defined in Eq. (12), which combines a T-F domain MSE loss and a time-domain SNR loss, we conducted an ablation study comparing it with two alternative training objectives: MSE loss only and SNR loss only.

As shown in Table IV, the proposed joint loss consistently achieves the best overall performance across all metrics. The MSE-only model tends to preserve the spectral structure but produces audible temporal artifacts, whereas the SNR-only model retains temporal fidelity at the expense of mild spectral distortion. By jointly optimizing both objectives, the network attains a better balance between spectral accuracy and perceptual quality, producing clearer and more natural-sounding enhanced speech. Moreover, combining complementary cues from both domains improves robustness and generalization under dynamic acoustic conditions.

TABLE IV
AVERAGE PESQ, STOI, AND DNSMOS FOR PROPOSED VI-NBFNET TRAINED WITH THE PROPOSED JOINT LOSS, MSE LOSS ONLY, AND SNR LOSS ONLY ACROSS SIRs OF -10, -5, 0, 5, 10, AND 15dB.

|  | Moving target speaker | | | | |
|---|---|---|---|---|---|
|  | PESQ | STOI | BAK | SIG | OVRL |
| Noisy | 1.334 | 0.656 | 1.383 | 1.699 | 1.303 |
| Proposed joint loss | **2.026** | **0.849** | 3.749 | 2.569 | **2.183** |
| MSE loss only | 2.022 | 0.849 | 3.714 | 2.390 | 1.987 |
| SNR loss only | 1.975 | 0.847 | **3.944** | 2.492 | 2.122 |

4) Visualization of attention weights: Fig. 10 visualizes the learned attention weights from the SAD during the estimation of time-varying SCMs in a moving-speaker scenario, providing an intuitive illustration of how the attention mechanism operates and dynamically responds to changes in the target speaker's location.

As shown in Fig. 10(a), the attention weights for speech SCM computation exhibit strong diagonal energy concentrations corresponding to speech-active segments. This indicates that the model effectively emphasizes frames surrounding the active speech regions, thereby maintaining temporal consistency and continuity in the estimated speech

SCMs. In contrast, for noise SCM computation shown in Fig. 10(b), the attention module mainly focuses on speech-inactive regions, such as the onset and offset of utterances. This observation reveals that the proposed attention mechanism dynamically adjusts its focus between speech-active and speech-inactive segments, enabling accurate estimation of both speech and noise SCMs under dynamic conditions.

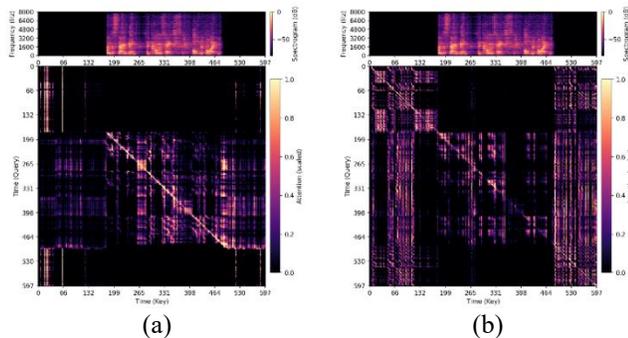

Fig. 10. Visualization of attention weights from the SAD in a moving-speaker scenario. (a) Attention weights for speech SCM computation, (b) attention weights for noise SCM computation.

5) Visual degradation robustness evaluation: The robustness of the proposed VI-NBFNet under challenging visual conditions was examined by simulating three types of visual degradation commonly encountered in practical audiovisual scenarios: partial occlusion, mosaic occlusion, and low-resolution degradation. In the partial occlusion condition, a black rectangular patch was placed over the mouth region to emulate face-mask wearing or partial obstruction. In the mosaic occlusion condition, the mouth area was pixelated to represent visually distorted inputs. In the low-resolution condition, each video frame was downsampled to 30% of its original width and height and then upsampled back to the original size, resulting in a blurred lip region that represents low-quality visual input.

Representative examples and the corresponding enhanced spectrograms are shown in Fig. 11, while the quantitative results are summarized in Table V. Despite the degradation in visual quality, VI-NBFNet maintained stable performance, with the differences across all tested conditions remaining below 0.05 for PESQ/STOI and below 0.1 for OVRL. These results indicate that the proposed audiovisual fusion strategy is highly resilient to moderate visual distortions, since the model primarily exploits temporally consistent lip-motion cues rather than fine-grained pixel-level details. Furthermore, as illustrated in Fig. 11, the enhanced spectrograms corresponding to occluded, mosaic, and low-resolution videos still preserve the harmonic structure and formant trajectories of the target speech, confirming that VI-NBFNet effectively generalizes to visually degraded inputs without significant performance loss.

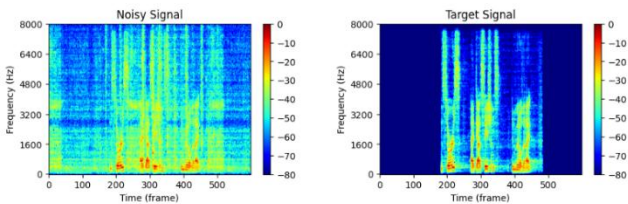

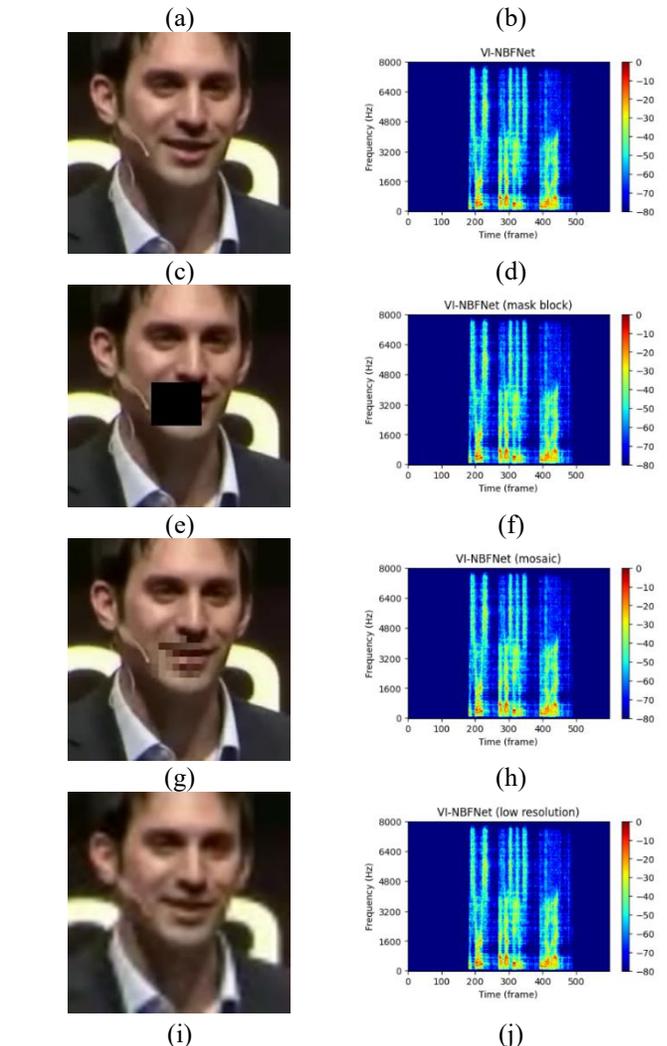

Fig. 11. Robustness evaluation of VI-NBFNet under visually degraded conditions. (a) Noisy input spectrogram, (b) clean target speech spectrogram, (c) original video frame, and (d) enhanced spectrogram produced by VI-NBFNet, (e) video with partial mouth occlusion, and (f) its enhanced spectrogram, (g) video with mosaic occlusion, and (h) its enhanced spectrogram, (i) low-resolution video input, and (j) its enhanced spectrogram.

TABLE V
A PERFORMANCE COMPARISON OF THE VI-NBFNET MODEL UNDER VISUALLY DEGRADED DATA CONDITIONS, INCLUDING MASK BLOCKING, MOSAIC, AND LOW RESOLUTION.

|  | PESQ | STOI | BAK | SIG | OVRL |
| --- | --- | --- | --- | --- | --- |
| Noisy | 1.198 | 0.771 | 1.424 | 2.137 | 1.414 |
| Unimpaired | **2.184** | **0.957** | **4.007** | **2.850** | **2.624** |
| Mask blocking | 2.167 | 0.957 | 3.928 | 2.807 | 2.549 |
| Mosaic | 2.213 | 0.958 | 3.974 | 2.843 | 2.603 |
| Low resolution | 2.167 | 0.958 | 3.974 | 2.843 | 2.603 |

*D. Experiment Result*

In addition to the preceding simulations, experiments using real-data were conducted. As described in Section IV.A, the target speaker signal used in the recorded evaluation was generated using two methods: playback through a loudspeaker and live speech from a human speaker. The loudspeaker played back pre-recorded clean utterances from the simulated test set, which is crucial for synchronization between the audio and video signals. The live speaker is intended for a more realistic scenario, allowing us to ensure audiovisual synchronization through manual post-editing and slight head movements. The spectrograms employing loudspeaker reproduction and live speaker recording are depicted in Figs. 12 and 13, respectively. The mean values of the evaluation metrics for all experiments are summarized in Table VI. Note that only non-intrusive DNSMOS metrics were utilized for the evaluation process, given the unavailability of ground-truth clean speech for the live speaker scenario.

TABLE VI
EVALUATION USING DNSMOS FOR REAL-WORLD RECORDINGS.

|  | BAK | SIG | OVRL |
|---|---|---|---|
| Noisy | 1.904 | 1.382 | 1.378 |
| VI-SSE | 2.592 | 2.513 | 1.972 |
| VI-MSE (BF) | 1.761 | 2.254 | 1.732 |
| VI-MSE (BF+PF) | 2.567 | 2.679 | 1.952 |
| VI-SA-BF | 2.109 | 1.972 | 1.672 |
| VI-NBFNet | **2.743** | **2.829** | **2.193** |

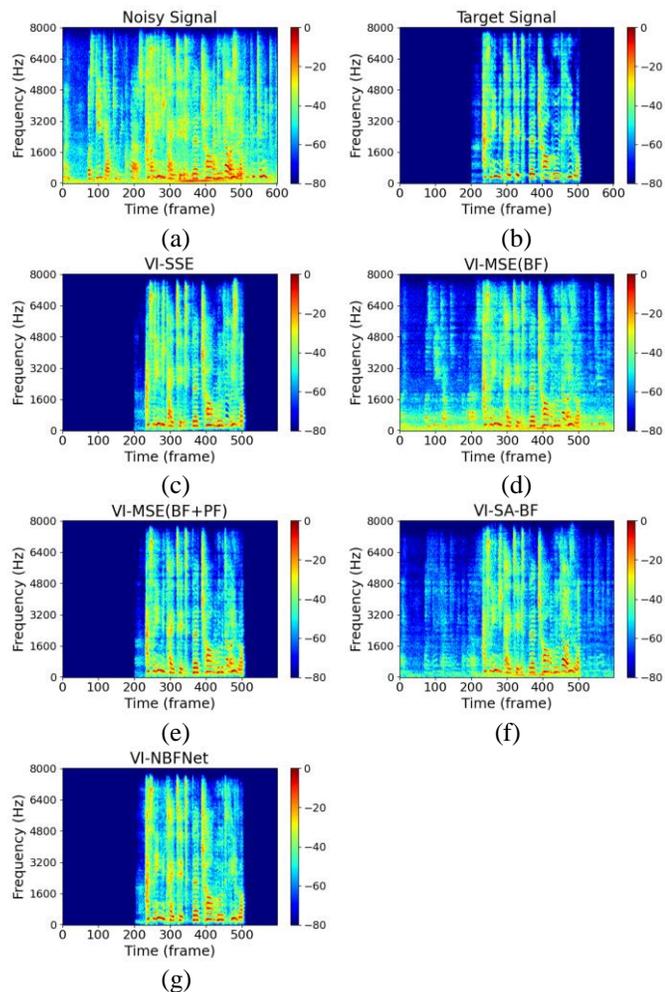

Fig. 12. Comparison of spectrograms of live recordings for the loudspeaker played back target speech. (a) Noisy signal, (b) clean target speech signal, (c) VI-SSE enhanced signal, (d) VI-MSE enhanced signal with beamforming, (e) VI-MSE enhanced signal with beamforming and postfiltering, (f) VI-SA-BF enhanced signal, and (g) VI-NBFNet enhanced signal.

As shown in Table VI, VI-NBFNet attains the highest DNSMOS among all evaluated methods. Specifically, the notably high OVRL score suggests that VI-NBFNet achieves a superior balance between speech quality and noise reduction in comparison to other approaches. Additionally, the SIG scores obtained using VI-SSE are lower than those of VI-MSE (BF+PF) and VI-NBFNet, which is consistent with the findings of simulation results. This finding further supports the assertion that single-channel approaches often fall short in terms of preserving speech quality. In the context of actual recording settings, VI-NBFNet attains higher SIG scores compared to VI-MSE (BF+PF). This result contrasts with the observed trend in simulation results. This finding indicates that the proposed method demonstrates superior enhancement performance and robustness under realistic conditions.

TABLE VII
EVALUATION USING WER (%) FROM WHISPER BASE AND TURBO ASR MODELS FOR REAL-WORLD RECORDINGS.

|  | base | turbo |
|---|---|---|
| VI-SSE | 93% | 27% |
| VI-MSE (BF) | 113% | 33% |
| VI-MSE (BF+PF) | 58% | 13% |
| VI-SA-BF | 75% | 35% |
| VI-NBFNet | **50%** | **8%** |

To further assess speech intelligibility in the absence of ground-truth clean speech, we additionally evaluated the Word Error Rate (WER) using two representative Automatic Speech Recognition (ASR) models from the Whisper family [57]: Whisper-base and Whisper-turbo, corresponding to medium- and large-capacity configurations, respectively. The turbo model serves as an optimized large-scale variant designed for high inference efficiency. As summarized in Table VII, the proposed VI-NBFNet achieves the lowest WER among all compared systems, yielding 50 % with Whisper-base and 8 % with Whisper-turbo. The considerable improvement from the base to the turbo model suggests that the enhanced speech produced by VI-NBFNet preserves clear phonetic content, enabling higher-capacity ASR models to better exploit its acoustic characteristics. Together with the DNSMOS results in Table VI, where VI-NBFNet achieved the highest scores across all metrics, the WER analysis further confirms that the proposed model yields highly intelligible and perceptually natural speech in realistic recording environments.

As expected, Fig. 13 shows poorer performance when using live-recorded video compared to simulated lip video from loudspeaker playback (Fig. 12), due to differences in visual training data. During training, lip imagery was selected from the LRS3 video set by clipping fixed-duration segments from each utterance. It is noteworthy that a specialized treatment is applied to the processing of facial images during speech pauses. Specifically, the initial and final video frames are utilized repeatedly to fill the periods where speech is absent. This approach ensures audiovisual continuity and synchronization during speech-absent periods. This is imperative for the audiovisual network to accurately associate lip movements with speech-related information for model training and inference. On the contrary, live recordings contain continuous facial movements even during speech silent periods, which the model was not trained for. As illustrated in Fig. 13, the recorded face image is smaller than those in the training set. This results in lower image resolution in the cropped lip image and performance degradation of the pretrained AVSR model. Another factor contributing to the performance degradation is the presence of uncontrollable environmental disturbances during recording, such as passing vehicle noise. However, the model is applicable for scenarios involving two directional sources. Even though there are challenges, VI-NBFNet is still the only method that can effectively suppress the non-target speech in Fig. 13. This shows that it is more robust and generalizable in real-world environments.

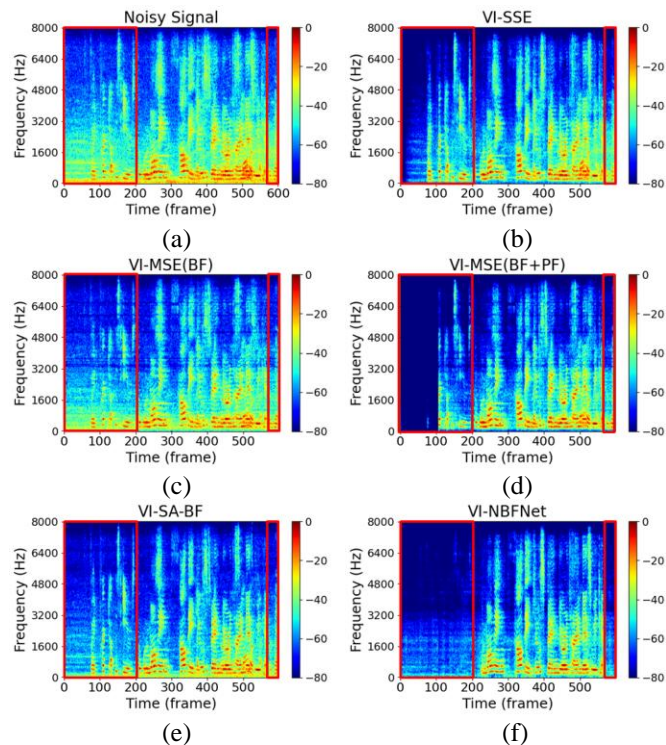

Fig. 13. Comparison of spectrograms of live recordings for the live-recorded target speech. (a) Noisy signal, (b) VI-SSE enhanced signal, (c) VI-MSE enhanced signal with beamforming, (d) VI-MSE enhanced signal with beamforming and postfiltering, (e) VI-SA-BF enhanced signal, (f) VI-NBF enhanced signal.

*E. Listening Test*

In addition to the aforementioned objective evaluation, a subjective evaluation of the proposed VI-NBFNet versus baselines was conducted using a subjective listening test based on the Multiple Stimuli with Hidden Reference and Anchor (MUSHRA) procedure [58]. A total of 21 participants took part in the subjective evaluation. The assessment of the perceived SE performance entailed a rating of four perceptual attributes by the participants. The following attributes were considered: Noise Suppression Level, Speech Intelligibility, Signal Distortion, and Overall Quality. The Noise Suppression Level was used to assess the extent to which interference was reduced, while the Speech Intelligibility Index was employed to evaluate the clarity and comprehensibility of the enhanced speech. The evaluation of audio artifacts introduced during the enhancement process was conducted through Signal Distortion index. The participants' perceptual preferences, as reflected in the Overall Quality, were compared to the reference signal. The attributes were rated on a continuous scale ranging from 1 to 100. To ensure the reliability of the results, participant responses were excluded if the hidden reference was rated below 90 or the anchor was rated above 10.

In each test case, participants were presented with four processed signals, including the proposed method VI-NBFNet and three baseline methods: VI-SSE, VI-MSE (BF+PF), and VI-SA-BF. A hidden reference corresponding to the ground-truth clean signal was also included. Furthermore, the lowpass-filtered reference signals at cutoff frequencies of 2 kHz and 4 kHz were utilized as anchors. To ensure a comprehensive evaluation, the listening test incorporated three realistic interference signals: competing speaker, music, and copy machine noise. These signals were also employed during model training. These cases were chosen to represent three common types of noise: speech-like interference, music signals, and background machine noise.

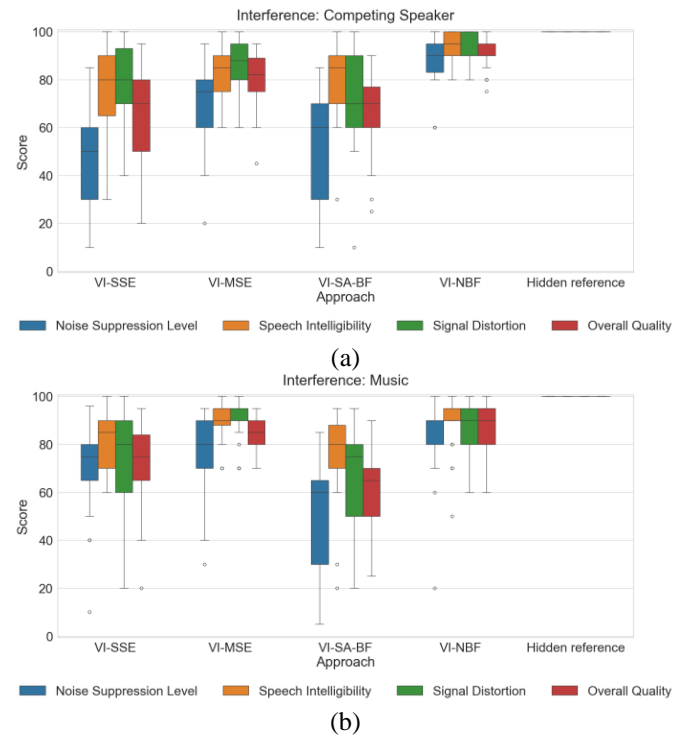

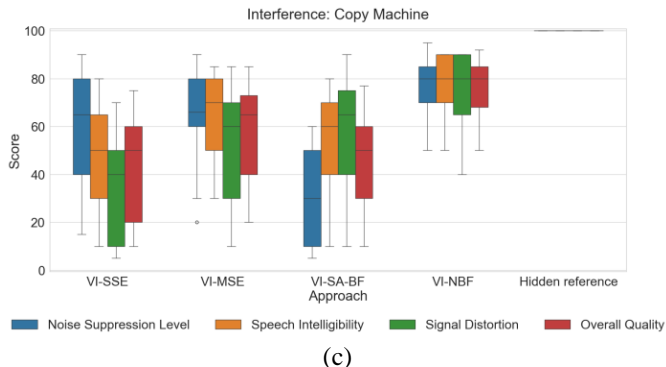

(c)

Fig. 14. The boxplots of listening test results for (a) a competing speaker, (b) music, and (c) a copy machine. The central line in each box represents the median, while the lower and upper bounds of the box indicate the 25th (Q1) and 75th (Q3) percentiles, respectively. The whiskers extend to the maximum and minimum values within the non-outlier range, while any outliers are represented as individual points beyond these bounds.

To assess the statistical significance of the differences in the results among the enhancement methods in the listening test, a series of statistical analyses were conducted. First, the normality of the ratings for each method under each interference condition (competing speaker, music, and copy machine) was evaluated using the Shapiro–Wilk test [59]. This analysis revealed that all p-values exceeded 0.05, thereby indicating that the normality assumption was met across all groups. Next, Levene's test [60] was used to examine the homogeneity of variances. For the competing speaker and music conditions, p-values were below 0.05, suggesting that the assumption of equal variances was violated. Therefore, Welch's Analysis of Variance (ANOVA) [61] was conducted. The results showed a significant difference among the enhancement methods for both conditions: $F(3, 40.35) = 23.11$, $p < 0.001$ for competing speaker, and $F(3, 42.34) = 12.55$, $p < 0.001$ for music. In contrast, for the copy machine case, Levene's test yielded a p-value above 0.05, confirming the homogeneity of variances. A standard one-way ANOVA [62] was thus performed to show a significant difference among the methods: $F(4, 100) = 41.45$, $p < 0.001$. Given the significant statistical difference in all three conditions, post hoc pairwise comparisons were subsequently conducted and the results are summarized in Table VIII.

TABLE VIII
THE P-VALUES OF THE OVERALL QUALITY FOR THE (A) COMPETING SPEAKER, (B) MUSIC, AND (C) COPY MACHINE NOISE. A P-VALUES LESS THAN 0.05 INDICATES HIGH STATISTICAL DIFFERENCE.

(a)

|  | VI-SSE | VI-MSE (BF+PF) | VI-SA-BF | VI-NBFNet |
|---|---|---|---|---|
| VI-SSE |  | 0.017 | 1.000 | 0.000 |
| VI-MSE (BF+PF) |  |  | 0.036 | 0.005 |
| VI-SA-BF |  |  |  | 0.000 |
| VI-NBFNet |  |  |  |  |

(b)

|  | VI-SSE | VI-MSE (BF+PF) | VI-SA-BF | VI-NBFNet |
|---|---|---|---|---|
| VI-SSE |  | 0.024 | 0.799 | 0.009 |
| VI-MSE (BF+PF) |  |  | 0.000 | 1.000 |
| VI-SA-BF |  |  |  | 0.000 |
| VI-NBFNet |  |  |  |  |

(c)

|  | VI-SSE | VI-MSE (BF+PF) | VI-SA-BF | VI-NBFNet |
|---|---|---|---|---|
| VI-SSE |  | 0.022 | 0829 | 0.000 |
| VI-MSE (BF+PF) |  |  | 0.169 | 0.024 |
| VI-SA-BF |  |  |  | 0.000 |
| VI-NBFNet |  |  |  |  |

The results of the listening test are presented as boxplots in Fig. 14. Across all three types of interference, the proposed VI-NBFNet demonstrated superior performance in most subjective attributes when compared to the baseline methods. It is noteworthy that VI-NBFNet attained the highest median score in Overall Quality for all conditions, with narrower box widths indicating more consistent ratings by the listeners. In particular, under conditions of competing speaker and copy machine noise, VI-NBFNet demonstrated significant superiority over all baseline methods, including VI-SSE, VI-MSE, and VI-SA-BF, with p-values less than 0.05 in all pairwise comparisons. In the context of music interference, VI-NBFNet exhibited superior performance in comparison to VI-SA-BF and VI-SSE (p-values < 0.05). However, the results did not reach statistical significance (p-value = 1), in contrast to those of VI-MSE. This outcome can be attributed to the relatively high SIR in the scenario, which has the effect of rendering VI-MSE perceptually comparable to VI-NBFNet. In all noise conditions, the single-channel method VI-SSE consistently received the lowest ratings, which is likely due to greater perceptual distortion. These findings underscore the efficacy and performance enhancement achieved by integrating spatial cues provided by array beamforming with spectral-temporal cues in SE without compromising audio quality.

## V. CONCLUSION

In this study, we proposed a novel visual-informed speech enhancement network VI-NBFNet that integrates attention-based beamforming into an end-to-end learning architecture. In contrast to conventional two-stage pipelines, which estimate masks and beamforming weights separately, VI-NBFNet learns both components within a unified framework. This approach results in effective interference suppression. The proposed system integrates the spatial and audiovisual modalities with an attention mechanism for SCM estimation, achieving superior beamforming performance, even in dynamic acoustic environments. A comparison of VI-NBFNet and the single-channel end-to-end system VI-SSE reveals that the former exhibits superior noise suppression with minimal speech distortion. In summary, the proposed method provides a solution for audiovisual speech enhancement that demonstrates excellent performance and robustness in both static and moving speaker scenarios. In the future, we will consider perceptually motivated loss functions such as SI-SDR or DNSMOS-based losses to better align the training objective with human hearing. The training data is being expanded to include a greater range of diverse and realistic conditions, such as mixtures with more than two speakers or moving interferers, which is expected to further improve the generalization of the proposed system.